\documentclass[aps,prb,twocolumn,superscriptaddress]{revtex4-1}
\usepackage{graphicx}
\usepackage{color}
\usepackage{dcolumn}
\usepackage{bm}
\begin{document}
\title{Electronic structure of graphene hexagonal flake subjected to triaxial stress}
\author{M. Neek-Amal, L. Covaci, Kh. Shakouri, F. M. Peeters }
\affiliation{Departement Fysica, Universiteit Antwerpen,
Groenenborgerlaan 171, B-2020 Antwerpen,
 Belgium.}
\date{\today}
\begin{abstract}
The electronic properties of a triaxially strained hexagonal
graphene flake with either armchair or zig-zag edges are investigated
using molecular dynamics simulations and tight-binding calculations.
 We found that: i) the pseudo-magnetic field  in the strained graphene flakes is
 not uniform neither in the center nor at the edge of zig-zag terminated flakes, ii) the
pseudo-magnetic field is almost zero in the center of armchair
terminated flakes but increases dramatically near the edges, iii)
the pseudo-magnetic field increases linearly with strain, for
strains lower than 15$\%$ while growing non-linearly beyond this
threshold, iv) the local density of states in the center of the
zig-zag hexagon exhibits pseudo-Landau levels with broken
sub-lattice symmetry in the zero'th pseudo-Landau level, and in
addition there is a shift in the Dirac cone due to strain induced
scalar potentials. This study provides a realistic model of the
electronic properties of inhomogeneously strained graphene where the
relaxation of the atomic positions is correctly included together
with strain induced modifications of the hopping terms up to next-nearest neighbors.
\end{abstract}



\maketitle

\section{Introduction}

  Strain engineering can be used to control the electronic properties
 of nanomaterials. This is of interest for fundamental physics, but is
 also relevant for potential device applications in nanoelectronics.
 Because the electronic and mechanical properties of an atomic monolayer of graphene
  are strongly influenced by strain they have attracted considerable attention over the last
years~\cite{gomes_designer_2012,pereira_tight-binding_2009,castro_neto_electronic_2009,guinea_generating_2010,
fogler_pseudomagnetic_2008,guinea_energy_2009,levy_strain-induced_2010,low_strain-induced_2010,vozmediano_gauge_2010,yamamoto_princess_2012,klimov_electromechanical_2012}.
Uniaxial strain was found to shift or even merge the Dirac
cones~\cite{goerbig_tilted_2008,montambaux_merging_2009,pereira_tight-binding_2009},
however, neither uniaxial nor isotropic strain produces a
pseudomagnetic field. Such fields can be produced only by
inhomogeneous strain which, when applied to graphene, alters the
hopping terms between nearest neighbors such that the additional
contribution can be seen as a pseudo-magnetic field with opposite
sign in the two Dirac cones, preserving therefore the time-reversal
symmetry~\cite{masir_pseudo_2013,guinea_generating_2010,guinea_energy_2009,low_strain-induced_2010,vozmediano_gauge_2010}.
The strain induced modifications of the electronic properties were
subsequently confirmed by STM measurements of highly strained
nanobubbles that form when graphene is grown on a Pt (111) surface
through the observation of Landau levels. These were shown to
correspond to strain-induced pseudo-magnetic fields larger than
300~T\cite{levy_strain-induced_2010}. Therefore, the study of
non-uniform strain distributions at the atomic scale is a promising
road for strain engineering
purposes~\cite{neek-amal_strain-engineered_2012,*neek-amal_strain-engineered_2012-1},
e.g., the possibility to generate a band gap. In the pioneering work
by Guinea \emph{et al.}~\cite{guinea_energy_2009} triaxial strain
applied to an hexagonal flake of graphene was shown to induce an
energy gap. This is a direct consequence of generating a constant
pseudo-magnetic field profile at the center of the
 flake while being variable only at the corners. Recent theoretical studies improved
 several aspects of the initial theory~\cite{kitt_lattice-corrected_2012,*kitt_erratum:_2013,de_juan_space_2012,linnik_effective_2012,masir_pseudo_2013,vozmediano_gauge_2010}.
 Most of these works concern the strain induced modifications of the continuum low energy
 Dirac Hamiltonian which become more complicated as additional orders in strain and momentum are included \cite{masir_pseudo_2013}.
 One such important correction is the spatial and strain dependent Fermi velocity~\cite{de_juan_space_2012,masir_pseudo_2013}. This shows that in addition to the pure gauge
 field given by the first order in strain, additional momentum dependent terms
 appear. It is also interesting to note that applying the triaxial
 stress on boron-nitride sheet decreases energy gap and localizes
 the frontier orbital in the center of hexagonal boron-nitride sheet~\cite{JPCP2013}.

By using an atomistic model, we choose both to express the gauge
fields in terms of their full tight-binding
expression~\cite{masir_pseudo_2013} and
 to describe the electronic properties directly from the tight-binding Hamiltonian with inhomogeneous
 hopping parameters calculated for the deformed graphene flake.

Furthermore, flakes of graphene can be useful for quantum dot
applications. There are several studies which address the latter in
terms of the energy spectrum of (unstrained) graphene flakes with
different sizes and different edge
structures~\cite{heiskanen_electronic_2008,*zhang_tuning_2008}, within both the
continuum model and the tight binding discrete model for the Dirac
equation~\cite{zarenia_energy_2011}.
 Recently it was demonstrated experimentally that by straining locally suspended graphene with an STM tip, it is possible to create such quantum dots~\cite{klimov_electromechanical_2012}.
Recently the resonant tunneling in graphene quantum dot
 was studied by Z. Qi~\emph{et al} using triaxially stressed graphene flake~\cite{qi_resonant_2013}.

In this study we analyze the effect of triaxial strain on the electronic properties of graphene by using an atomistic model which fully takes
 into account the relaxation of the graphene lattice using bond order interatomic potential.
 Based on our simulations and using tight-binding theory we show that several of the predictions
 of Ref.~[\onlinecite{guinea_energy_2009}]  should be modified resulting in new physical effects.

We found that the vector potential in a hexagonal flake of graphene
under triaxial strain behaves differently for zig-zag or armchair
terminated edges and the corresponding induced pseudo-magnetic field
is spatially inhomogeneous in both cases.  We also show that an
energy gap appears in the zig-zag hexagon upon applying triaxial
strain. This is due to the appearance of pseudo-Landau levels, which
form when the pseudo-magnetic field is large enough such that the
magnetic length is smaller than the flake size. For hexagons with
armchair edges we find that pseudo-Landau levels are absent, due to
the small induced pseudo-magnetic field in the center, and that the
local density of states is enhanced mainly at the edge of the
sample. 

The paper is organized as follows. In Sec.~II we review the theory of elasticity for triaxial strained graphene. In Sec. III we present
our molecular dynamics simulation method for applying triaxial strain on a hexagonal flake. In
Sec. IV, the tight binding model used for calculating the pseudo vector potential and pseudo magnetic field is introduced. Section V includes the main results of
our work, i.e. lattice deformation obtained from molecular dynamics simulation, vector potential and pseudo magnetic field and local density
of states for both zig-zag and armchair flakes.

\section{Elasticity theory for triaxial stress}

\begin{figure}
\begin{center}
\includegraphics[width=0.95\linewidth]{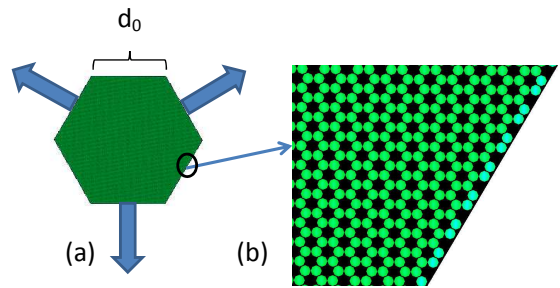}
\caption{(Color online) (a) Hexagonal un-strained flake of graphene with armchair edges.
 The arrows in (a) indicate the triaxial strain directions. (b) The zoomed part of one of the edges. \label{fig1}}
\end{center}
\end{figure}

Figure~\ref{fig1}(a) shows a hexagonal graphene flake having
armchair edges with side size $d_0$. The blue arrows indicate the
triaxial stress directions which are along the three equivalent
crystallographic directions. In Fig.~\ref{fig1}(b) the edge
structure is shown. In polar coordinates $(r,\theta)$ the applied
triaxial stress results in a displacement vector
$\vec{U}=(U_r,U_{\theta})=C\,r^2(\sin(3\theta),\cos(3\theta))$,
where $C$ is a constant determining the strength of the applied
stress which has the dimension of inverse
length\cite{guinea_energy_2009,vozmediano_gauge_2010}. The
displacement can be written in Cartesian coordinates as
\begin{equation}
\vec{U}=C(2x\,y,\,x^2-y^2).\label{EqUx}
\end{equation}

On the other hand linear elasticity theory for an isotropic material leads to the stress-strain relation
 $\sigma_{jk}=\lambda\, \delta_{jk}\,u_{jj}+2\mu\, u_{jk}$, where $\lambda$ and $\mu$ are the Lam\'{e} parameters that
determine the stiffness of a material and $u_{jk}$ are elements of the strain tensor. If we substitute $\vec{U}$ in Eq.~(\ref{EqUx})
the components of the stress tensor in cartesian coordinates can be found as
\begin{equation}
  \sigma(x,y)= 4\mu\,C\,\left(
  \begin{array}{cc}
   y & x \\
    x &  -y \\
  \end{array}
\right),\label{Eqsx}
\end{equation}
where the $x$ axis is taken along the zig-zag direction and the $y$ axis along the armchair direction.
 Here both zig-zag and armchair hexagonal graphene flakes are studied separately. Notice that the three stressed edges
 (and also the free edges) should have only one type of edge: zig-zag or armchair. The zig-zag hexagonal flake is not shown in Fig.~\ref{fig1}.
 In the following we present the methodology and compare our atomistic simulation results with those predicted by the above simple continuum elasticity theory.

\section{Molecular dynamics simulation with triaxial strain}

\begin{figure}
\begin{center}
\includegraphics[height=1.02\linewidth,angle=-90]{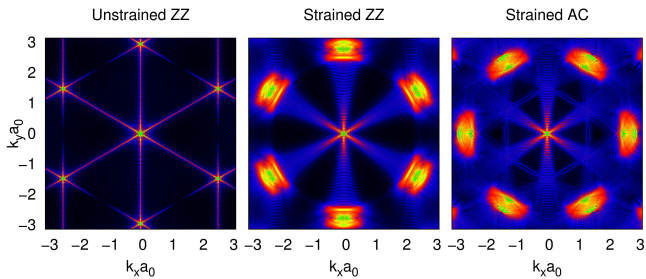}
\caption{(Color online)  Diffraction pattern for (a) un-deformed
hexagonal flake (b) zig-zag (c) and armchair hexagonal flakes under
triaxial stress. \label{fig2}}
\end{center}
\end{figure}

We apply triaxial stress on the edges of an hexagonal flake of
graphene as indicated in Fig.~\ref{fig1}. Two different samples with
armchair and zig-zag edges are considered and consist of 40234 and
40016 carbon atoms, respectively with optimized initial side
 i.e. $d_0=17.3$\,nm. When applying stress, all atoms in the three
none-alternative edges experience external constant force during
each molecular dynamics simulation (MD) which causes them to move
along the direction of the arrows shown in Fig.~\ref{fig1}(a). The
system reaches its equilibrium size under the applied force,
resulting in different strains for different forces. In our MD
simulations we have used the AIREBO
potentials\cite{stuart_reactive_2000} which is in particular
suitable for simulating hydrocarbons. The stretching process is done
at low temperature (namely T=10\,K). After reaching the desired
strain (i.e. we study strains up to the breaking point) we performed
energy minimization to find the minimum energy configuration using
the conjugate gradient method under the constant force condition. We
notice that there is no out-of-plane deformation in the final
minimized samples. We consider a measure of the strain determined by
$\epsilon=(d-d_0)/d_0$ where $d$ is the distance between the center
of the hexagonal flake and one of the edges under stress.  The final
strained samples are no longer perfect hexagons (see
Fig.~\ref{fig2}).

The applied stress changes the reciprocal lattice as well as the
real space lattice. Using the two dimensional Fourier transformation
of the final strained samples the change in the reciprocal lattice
(Brillouin zone) is obtained. Figures~\ref{fig2}(a,b,c) show the
diffraction pattern of the original un-strained, strained zig-zag
and armchair flakes, respectively where $\epsilon=13\%$. Notice that
the original K and K$^{\prime}$ valleys are altered differently due
to the in-plane triaxial strain. The variation in the diffraction
pattern is different from that of corrugated suspended graphene due
to intrinsic thermal ripples\cite{meyer_structure_2007}. It is
expected that such new patterns can be realized in experiment. The
direct lattice and the reciprocal lattice deformations alter the
hexagonal shape of the original Brillouin zone and will change the
vector potentials~\cite{kitt_lattice-corrected_2012,*kitt_erratum:_2013}.
\section{Tight binding model for the gauge field}
The electronic properties of graphene are described by a
 tight-binding Hamiltonian for the $\pi$ carbon orbitals.
We employ the Hamiltonian that describes the low-energy band
structure\cite{castro_neto_electronic_2009} ignoring the spin
degrees of freedom which is motivated by the fact that no
spin-flipping terms are present in the Hamiltonian. Strain is
included in the modified hopping amplitudes between the $\pi$
orbitals, $t_\pi(r_{ij})$, according to the empirical relation
$t_\pi(r_{ij})=t_0 e^{-\beta(|\vec{\delta}_{ij}|/{a_0}-1)}$, where
$\beta=-3.37$, $t_0=2.7eV$, $a_0=1.42$\AA~is the equilibrium
inter-carbon distance and $\vec{\delta}_{ij}$ is the vector which
connects the two neighboring atoms in the strained sample; here we consider both nearest and next-nearest neighbor terms. All neighbor distances are obtained
from the relaxed MD sample. Since the applied strain is in-plane the
effect of misalignment of the $\pi$ orbitals resulting from the
finite curvature is negligible.  More details of the used
tight-binding model and the related numerical techniques for
performing large system calculations can be found in our previous
works~\cite{covaci_superconducting_2011,covaci_efficient_2010,munoz_tight-binding_2012,*munoz_tight-binding_2013,neek-amal_strain-engineered_2012}.
\begin{figure}
\includegraphics[width=0.45\linewidth]{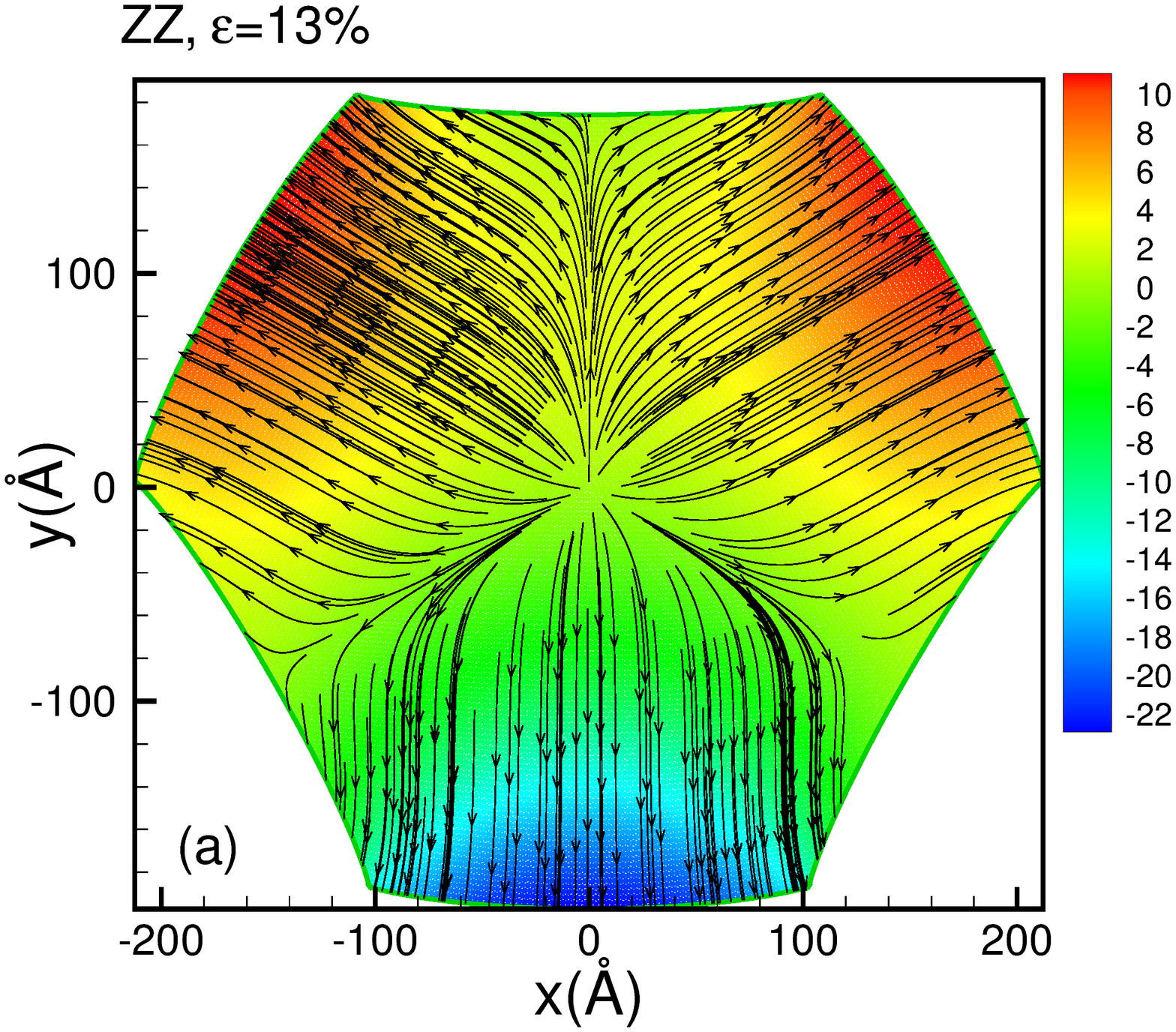}
\includegraphics[width=0.45\linewidth]{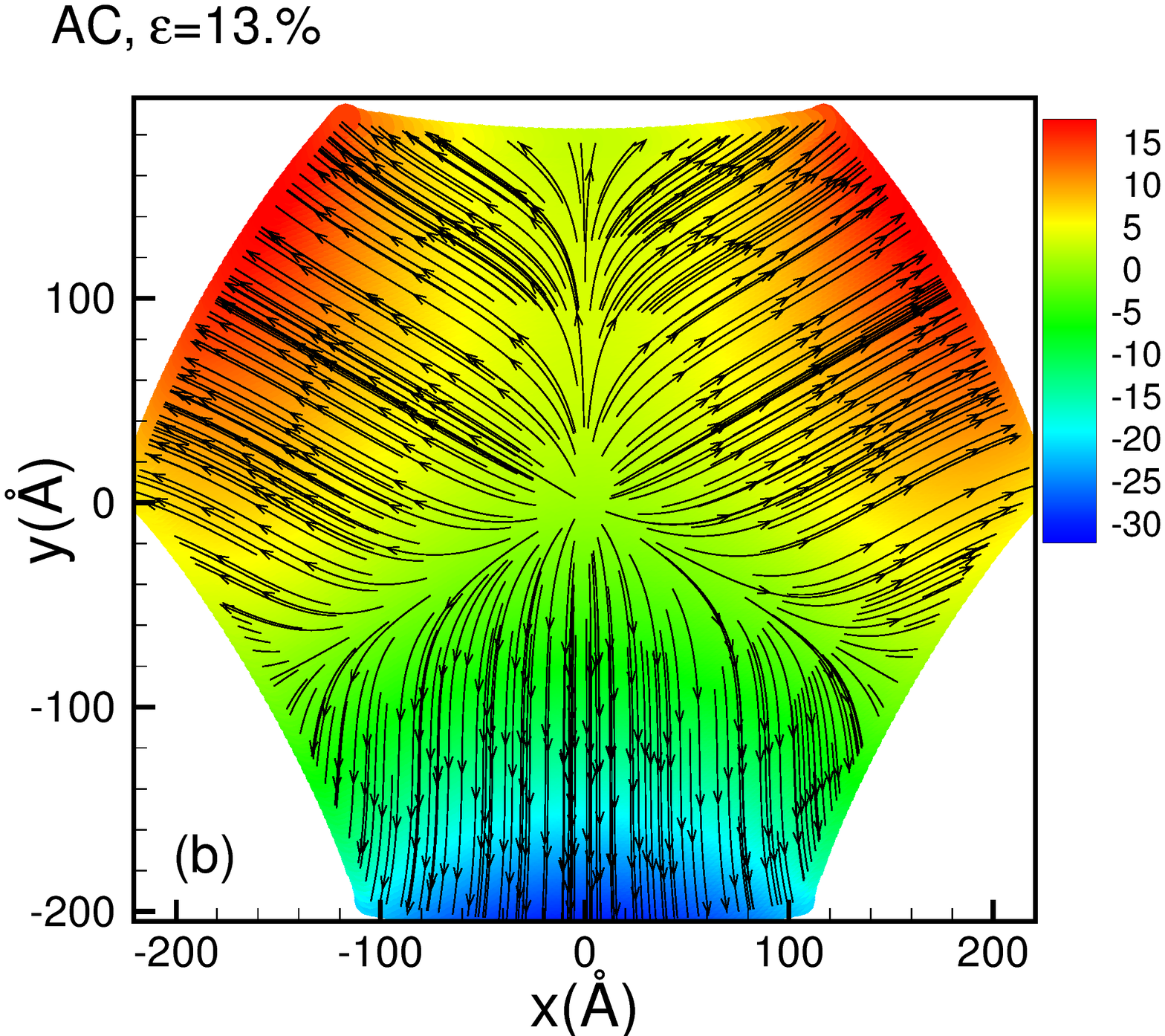}
\includegraphics[width=0.45\linewidth]{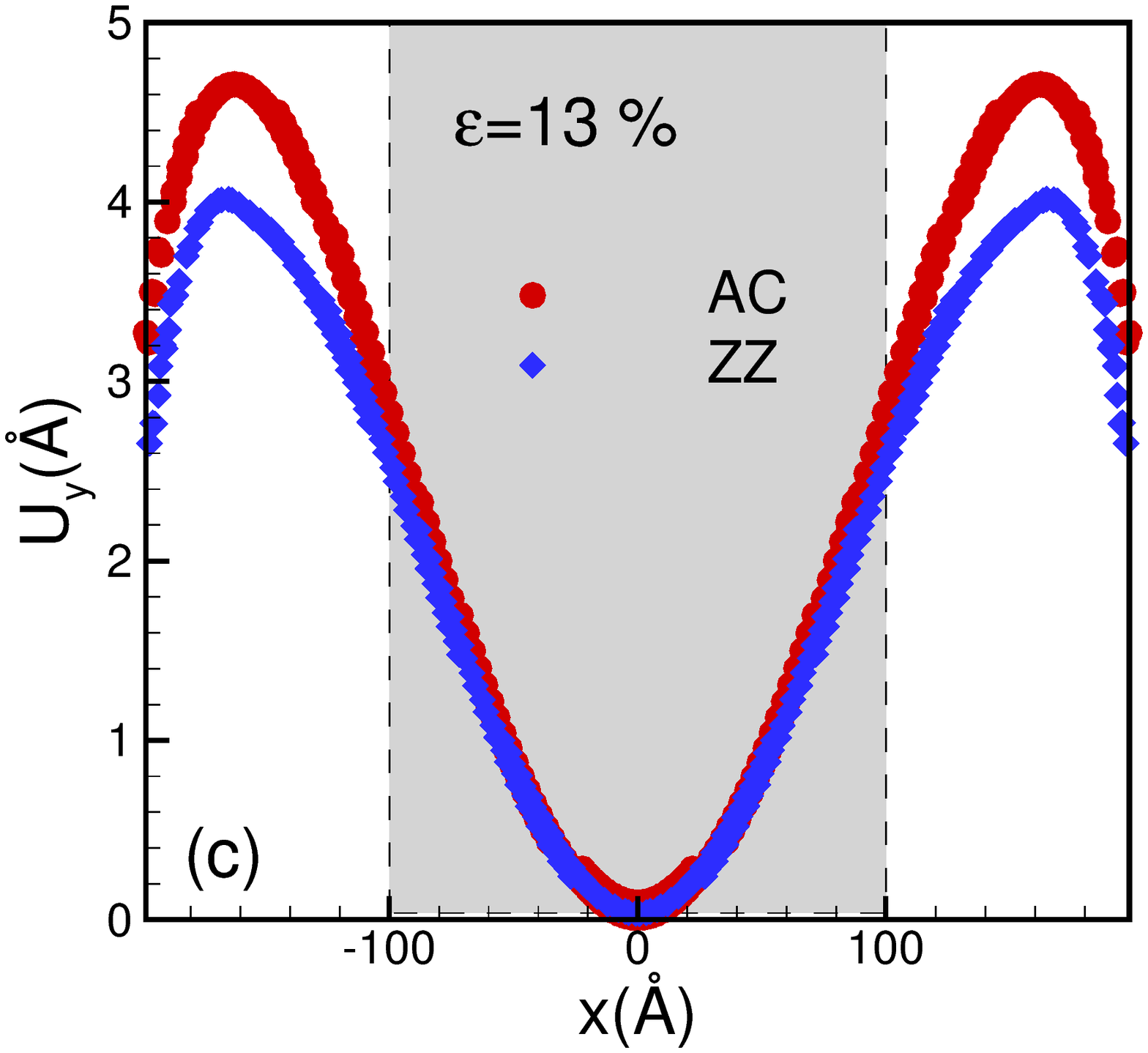}
\includegraphics[width=0.45\linewidth]{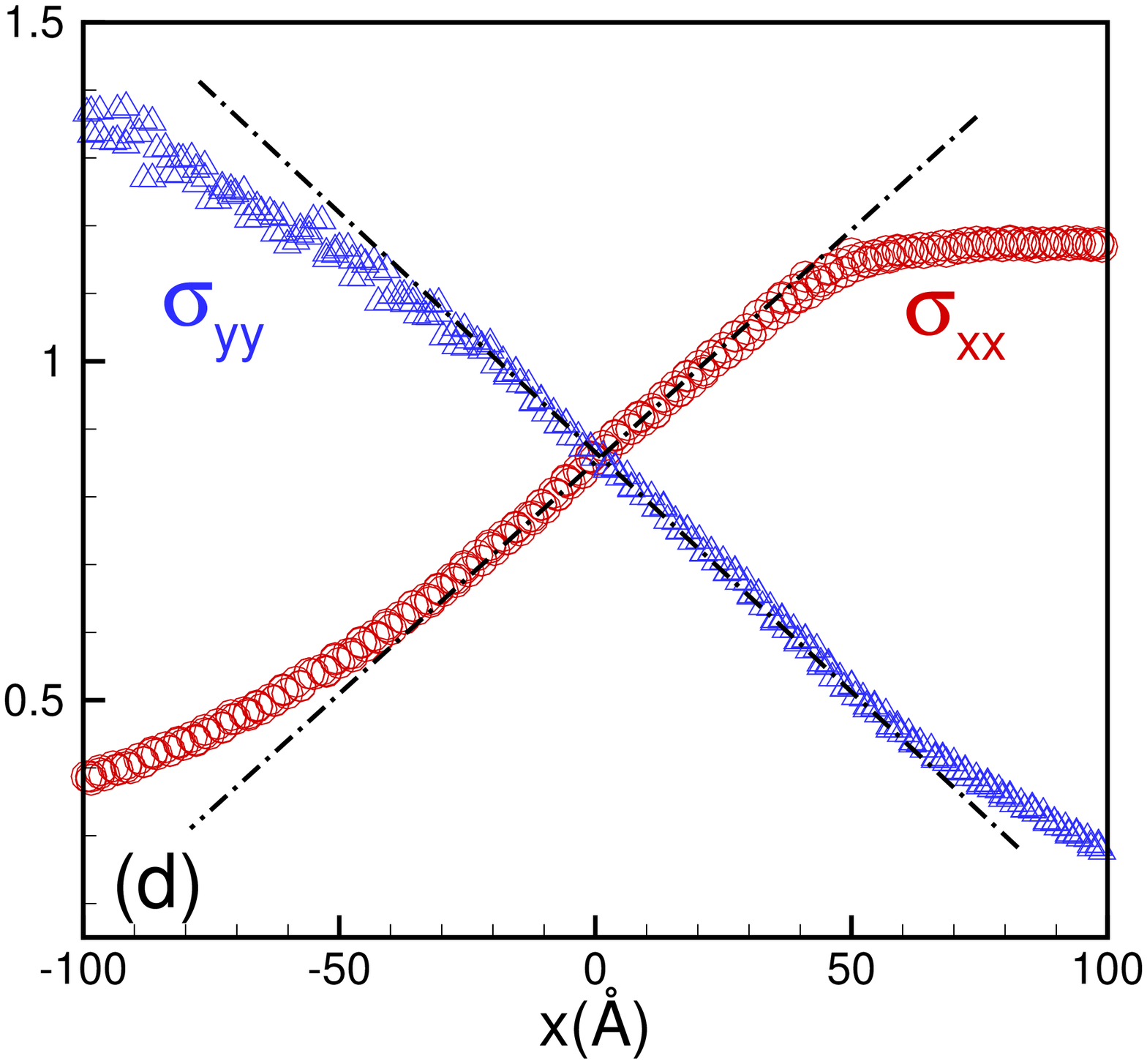}
\caption{(Color online) The displacement vector pattern in strained
(a) zig-zag (b) and arm-chair hexagonal graphene flakes. The
variation of (c) $U_y$ versus $x$ where $|y|<2\,$\AA~ for zig-zag
(circles) and armchair (triangles) hexagonal flakes subjected to an
inhomogeneous strain of about~$\epsilon=13\%$. In (d) the two
diagonal elements of the stress tensor, i.e. $\sigma_{xx}$ and
$\sigma_{yy}$ are compared to the prediction from elasticity theory
(dashed lines) where $|y|<2\,$\AA,~see the text. \label{fig3}}
\end{figure}
From  a theoretical point of view, external forces deform graphene
so that the nearest neighbor distances become non-equal and
 results in modified hopping parameters, which are now a
function of the atomic positions
$t(\textbf{r})$\cite{neek-amal_nanoengineered_2012,castro_neto_electronic_2009}. The Fermi surface is
displaced in reciprocal space ($\vec{k} \rightarrow
\vec{k}-\frac{e}{\hbar} \vec{A}$), where $\vec{A}$ is the fictitious
vector potential and $\vec k$ refers to the K-point) and consequently a
pseudo-magnetic field ($\vec{B}=\frac{1}{e\,v_F}{\vec{\nabla}\times
\vec{A}}$  where $e$ is the unit of charge and $v_F\sim10^6m/s$ is
the Fermi velocity in graphene\cite{castro_neto_electronic_2009})
appears\cite{levy_strain-induced_2010}. The new term should be added to the original
tight binding Hamiltonian due to the modification of the hopping
parameters which includes the induced gauge field:
\begin{equation}\label{gauge}
A_x+iA_y=\sum_{\vec{\delta}_{ab}} \delta t_{ab}(\vec{r}) e^{-i
\vec{k}.\vec{\delta}_{ab}},
\end{equation}
where $\delta t_{ab}=t-t_0$ is the difference between the hopping
parameters of the deformed and the original lattice. The position of
the $\vec{K}$ valley in the original lattice is
$\vec{K}=\frac{4\pi}{3\,a_0\sqrt{3}}(1,0)$ if the x-axis is taken
along the zig-zag direction. We use Eq.~(\ref{gauge}) to calculate
the pseudo-magnetic field which includes both changes in the hopping
parameters and lattice deformations. As recently shown in
Ref.~[\onlinecite{sloan_strain_2013}], this is the correct way to
extract the pseudo-magnetic field from atomistic simulations.

Notice that one can use the displacement vectors given by
Eq.~(\ref{EqUx}) for producing a deformed hexagonal flake without
molecular dynamics simulation relaxation. The latter can be used as
the input coordinates for calculating the gauge field and
the corresponding pseudo magnetic field. However we show in the appendix
that the resulting Landau levels and the obtained lattice
deformations are not consistent with those found here after fully
relaxation of the coordinates using the true relaxation mechanism of
the atomistic system.

\section{Results}

\subsection{Lattice deformation}

In Fig.~\ref{fig3} we show the lattice deformation due to the
triaxial strain (with $\epsilon=13\%$) in zig-zag terminated (a) and
armchair terminated (b) hexagonal flakes, respectively. The  arrows
indicate the displacement vector streamlines, i.e. the vector field
$\vec{U}$. In both cases the field refers to the triaxially stressed
systems. The colors indicate the direction of the displacement
vectors, i.e. red refers to  upward and blue refers to downward
displacements. The displacement vectors in zig-zag and armchair
flakes are similar but the electronic properties will be rather
 different. The streamline vectors are perpendicular to the
three stressed edges and are almost parallel to the free arc-shape
edges. The larger the strain the larger the concavity of the edges.

In order to compare our numerical results with those predicted by
Eq.~(\ref{EqUx}), we plot the $U_y$ components of the resulted
displacements from our MD simulations and the two main components of
the stress tensor in panels (c) and (d) of Fig.~\ref{fig3}. In
Fig.~\ref{fig3}(c) we show the variation of $U_y$ with $x$ for
constant $y$ (i.e. $|y|<2$\AA)~in the strained armchair (blue
diamonds) and zig-zag (red circles) which were subjected to the
strain $\epsilon=13\%$. There is  a clear deviation from linear
behavior (see $U$ in Eq.~(\ref{EqUx})) which is very pronounced
close to the borders of the sample. The same behavior is found for
$U_x$ but is not shown. In Fig.~\ref{fig3}(d) we show the variation
of $\sigma_{yy}$ and $\sigma_{xx}$ with $x$ for a typical strain
($13\%$). The lines indicate the stress component from
Eq.~(\ref{Eqsx}). It is again seen that there is considerable
deviation from the prediction of continuum elasticity theory
(Eq.~(\ref{EqUx})), in particularly beyond $|x|>$50\AA~ where the
zig-zag and armchair profiles deviate from each other; notice that
linear elasticity theory does not distinguish these two different
lattice orientations. The reason for these deviations is the effect
of the free edges on the lattice distortion which results in a
complex strain distribution. The latter effect was neglected in the
previous studies\cite{guinea_energy_2009,vozmediano_gauge_2010}.
Nevertheless, in most recent theoretical works the main attention
was directed to the center of the flake which as we see from
Figs.~\ref{fig3}(c,d), for the region $|x|<50\,$\AA, ~ agrees with
elasticity theory.

\subsection{Pseudo-magnetic field}
\begin{figure}
\includegraphics[width=0.48\columnwidth]{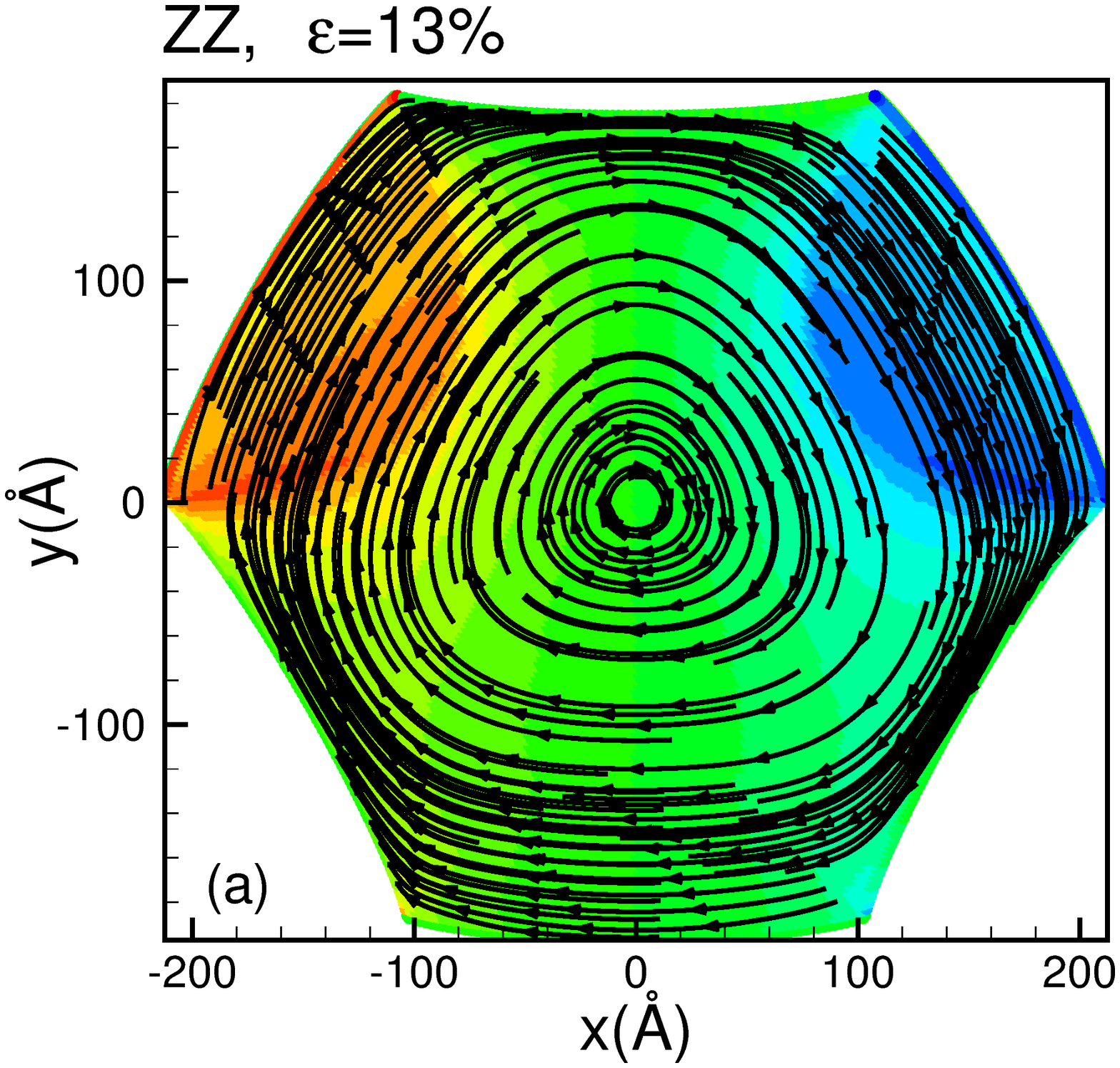}
\includegraphics[width=0.48\columnwidth]{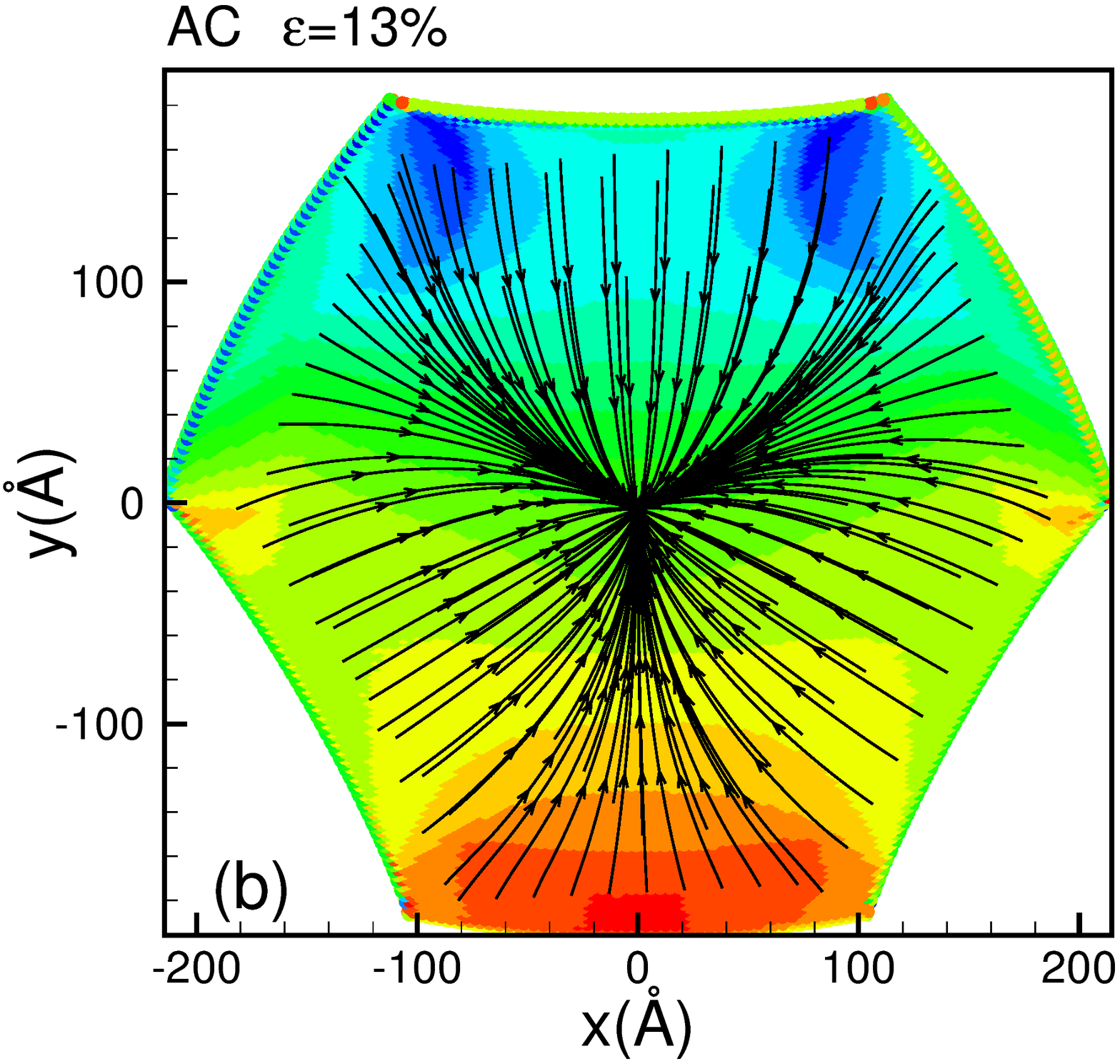}
\includegraphics[height=\columnwidth,angle=-90]{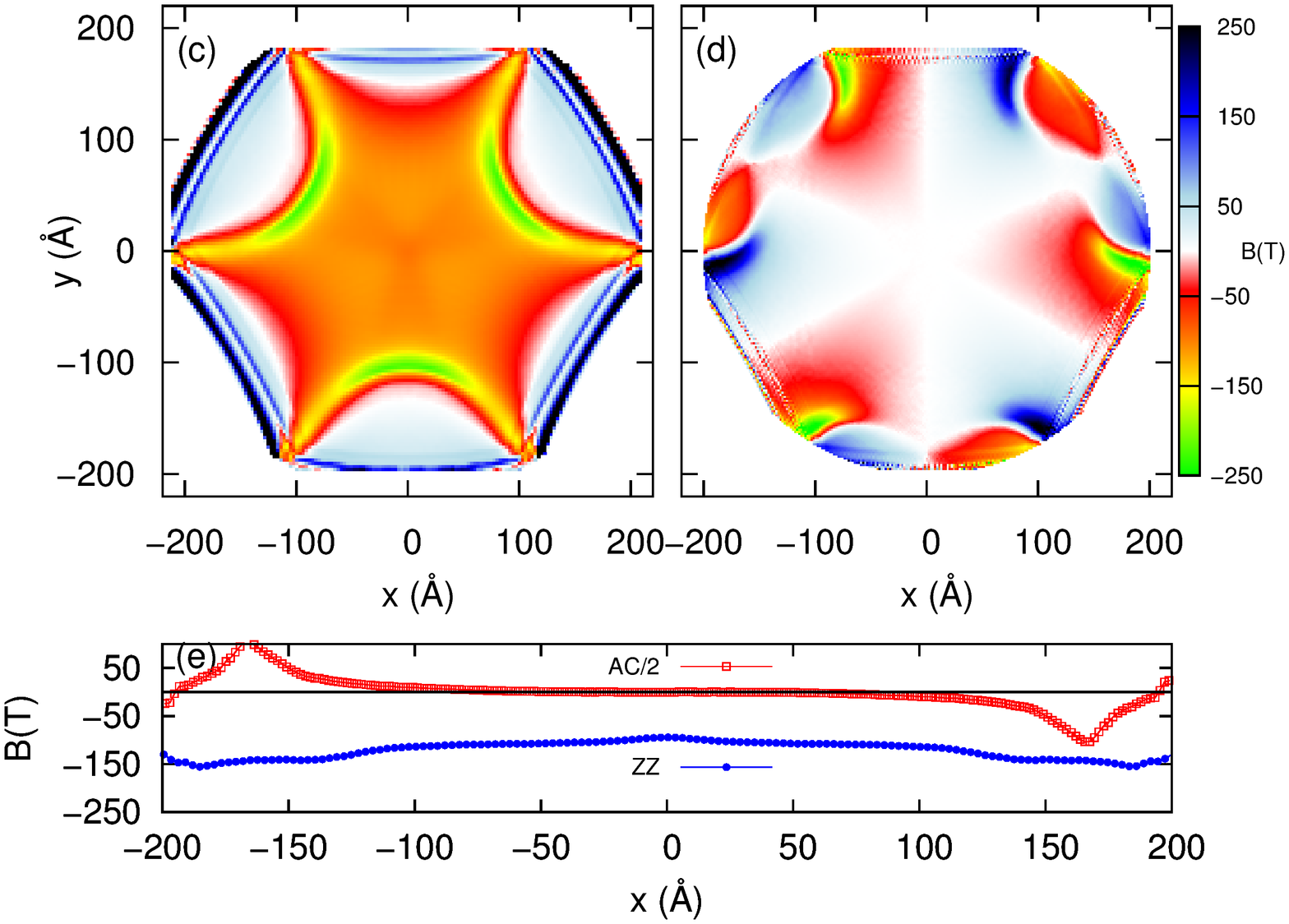}
\includegraphics[height=0.95\columnwidth,angle=-90]{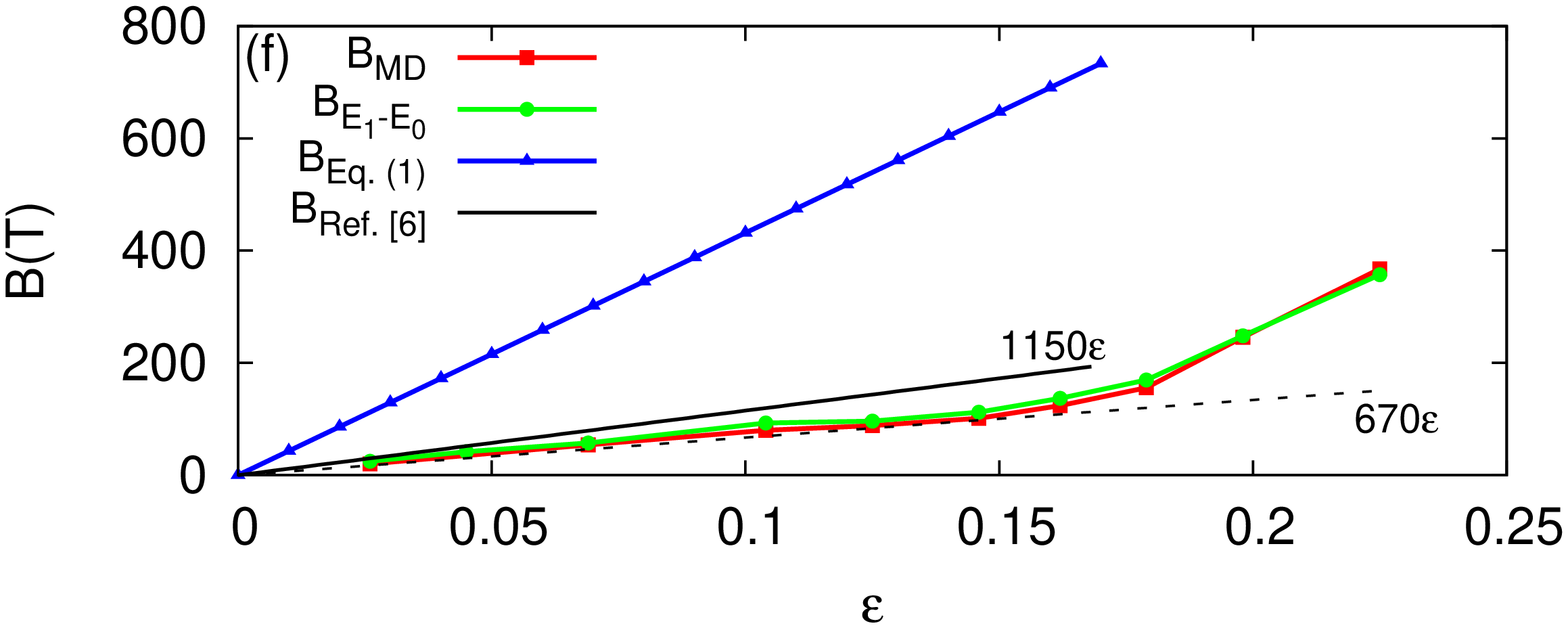}
\caption{(Color online)  Vector potential $\vec{A}$ for strained
hexagonal zig-zag (a) and armchair (b) graphene where
$\epsilon=13\%$. The corresponding pseudo-magnetic field is shown in
(c) for zig-zag and (d) armchair flakes. (e) The pseudo-magnetic
field profile along the $x$-axis for both armchair (red square) and
zig-zag (blue cricles) flakes. (f) Comparison between the
pseudo-magnetic field at the center of the zig-zag flake obtained
from the deformation (red squares), from the electronic gap between
the zero and first pseudo-Landau levels (green circles) and the
prediction from linear elasticity theory given by
Ref.~[\onlinecite{guinea_energy_2009}](blue triangles).
\label{fig4}}
\end{figure}
Using Eq.~(\ref{gauge}) we calculated the vector potential for  the
obtained lattice deformations from our MD simulations. It is interesting
that the vector potential streamlines exhibit neither constant nor
circular orbits. In the zig-zag flake the vector potential shows
orbits having deformed triangular shape (there is a kind of
 three fold symmetry, see Fig.~\ref{fig4}(a)) but surprisingly for the armchair flake they
do not exhibit orbits and the vectors follow an hyperbolic function,
Fig.~\ref{fig4}(b). Therefore the corresponding pseudo-magnetic
field for the two cases will be very different. The
 pseudo-magnetic field profiles as generated by the strain
configurations are shown in Figs.~\ref{fig4}(c) and~\ref{fig4}(d)
for the zig-zag and armchair flakes, respectively. The important
effect is the variation of the field over the zig-zag flake,
specially around the center of the flake~(see Appendix B). For the
armchair flake the induced magnetic field is close to zero and
varies smoothly in the central part, see Fig.~\ref{fig4}(d). For
better visualization we plot in Fig.~\ref{fig4}(e) the
pseudo-magnetic field along the arrows shown in
Figs.~\ref{fig4}(c,d). It is interesting to note that the
pseudo-magnetic field in the zig-zag flake exhibits three fold
symmetry while for the armchair it shows a more complex pattern. We
also present in Fig.~\ref{fig4}(f) a comparison between the central
pseudo-magnetic fields obtained directly from the deformation, the
field obtained from the electronic gap between the zeroth and first
pseudo-Landau levels and the prediction from
Ref.~[\onlinecite{guinea_energy_2009}], and that of resulted using
Eq.~(1)~,i.e. $B=\frac{16\,C\,\hbar}{a_0\,e}$. A linear regime is
found only for $\epsilon<15\%$, while beyond this value the
pseudo-magnetic field behaves non-linearly with respect to
$\epsilon$. The values obtained from the deformation and electronic
properties are very similar but much smaller than the prediction of
continuum elasticity theory (triangular symbols in
Fig.~\ref{fig4}(f)). We attribute this discrepancy to the change in
the lattice structure due to the relaxation of the graphene sheet.
We note that for the same $\epsilon$ the shape of the flake obtained
from the MD simulation is very different from the one obtained from
the deformation defined by Eq.~(\ref{EqUx}).

\subsection{Local density of states}
\begin{figure}
\begin{center}
\includegraphics[height=\linewidth,angle=-90]{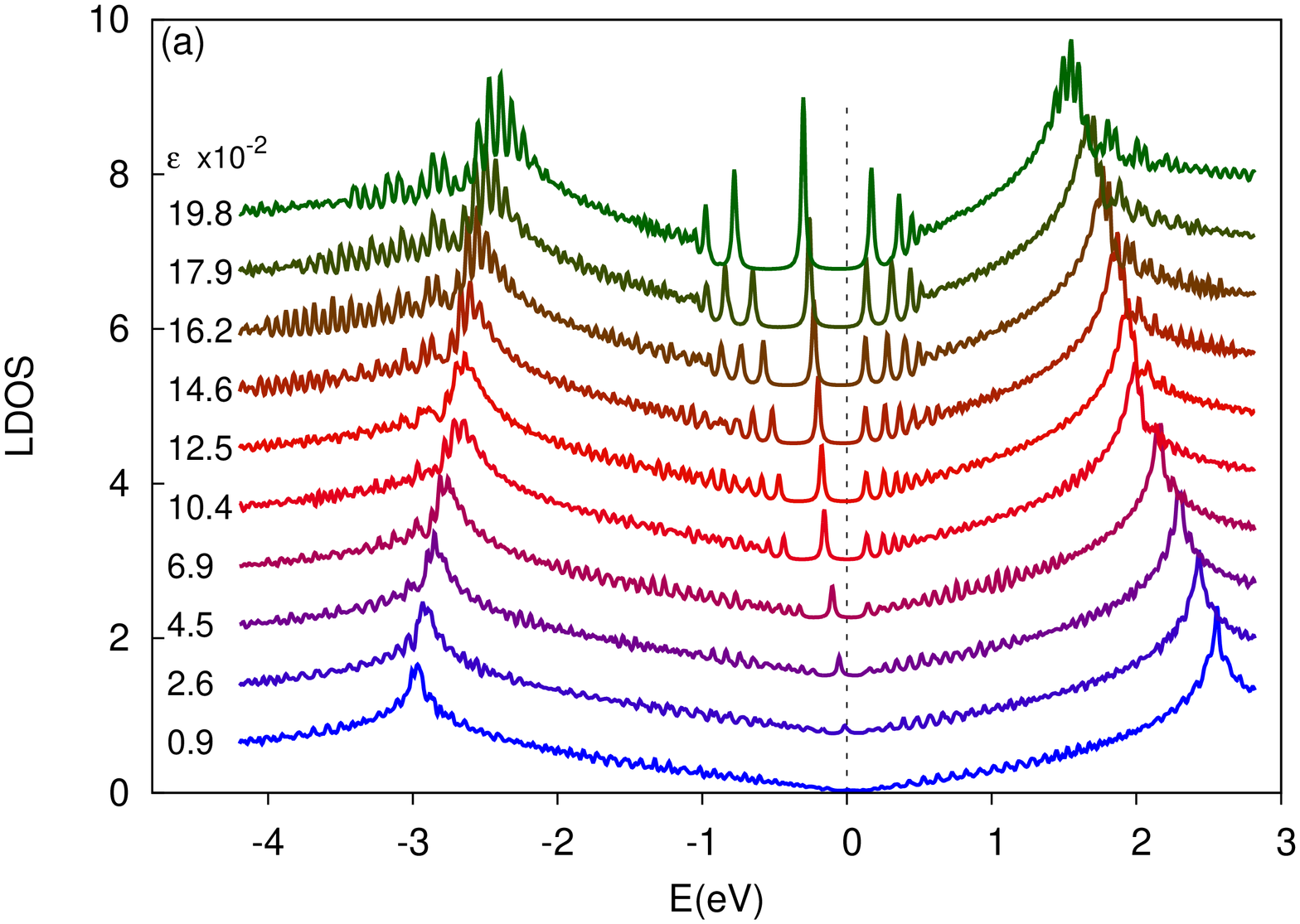}
\includegraphics[height=\linewidth,angle=-90]{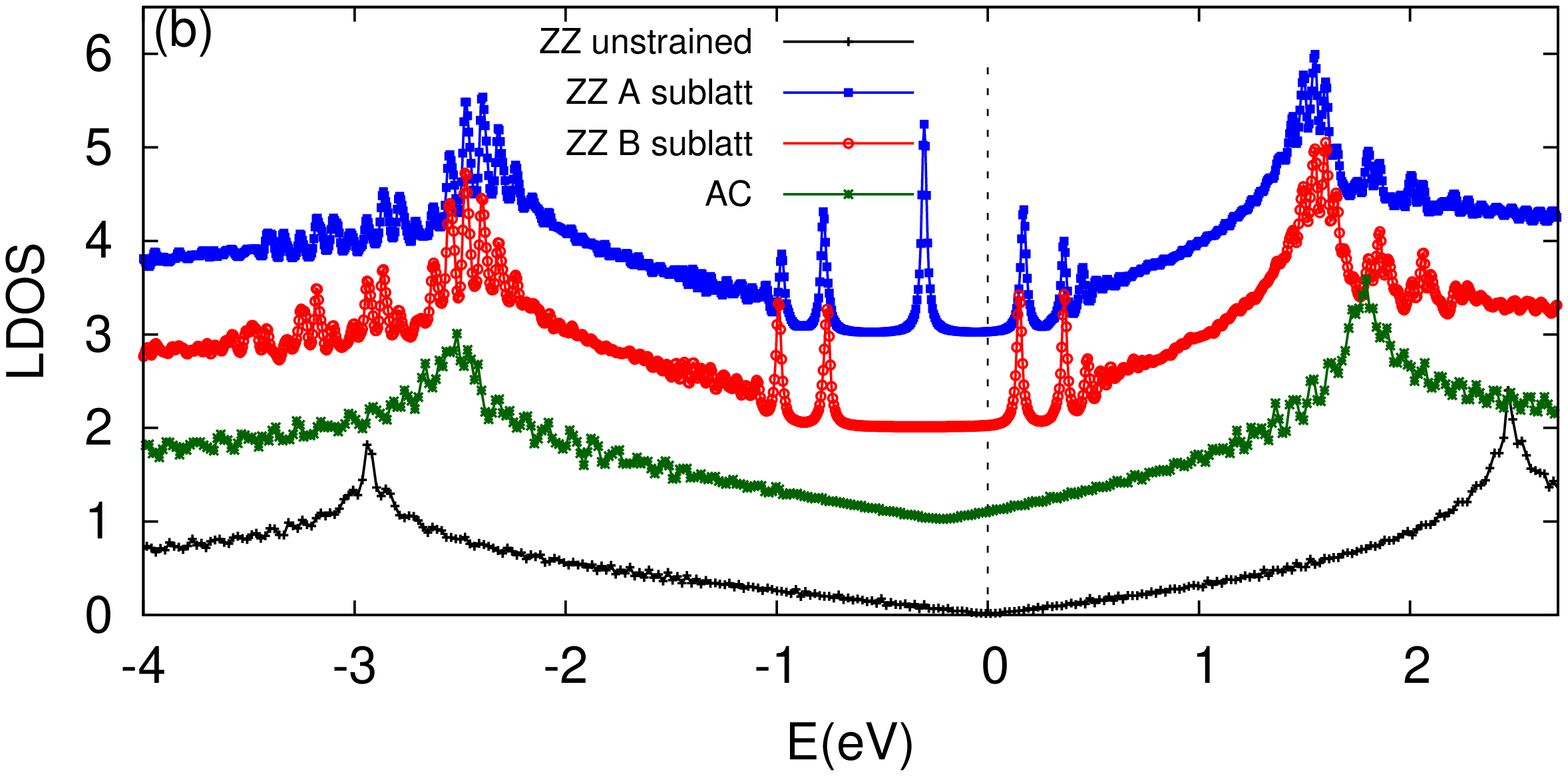}
\includegraphics[height=\linewidth,angle=-90]{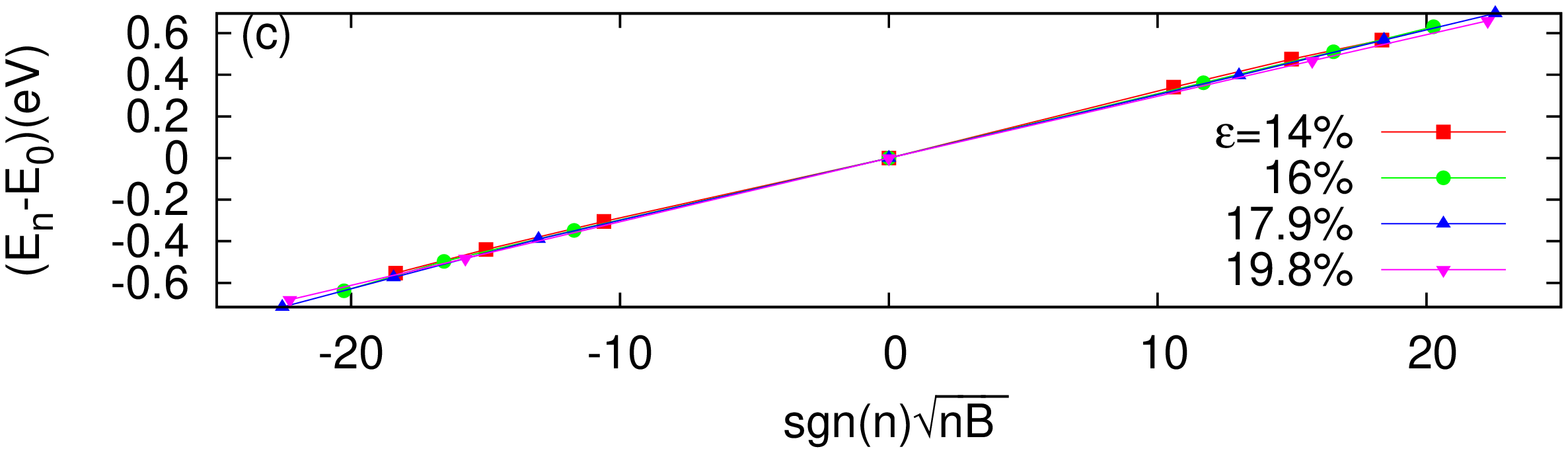}
\caption{(Color online) (a) Local density of states in the center of
the hexagon for the A-sublattice and various strains. (b) Comparison
between the central LDOS of the unstrained zig-zag flake, strained
zig-zag flake for both A and B sublattices and strained armchair
flake with strain of 13$\%$. (c) Plot of the positions of the
pseudo-Landau levels versus $sgn(n)\sqrt{nB}$, where $B$
 is extracted from the difference between the position of the zeroth and first Landau level. Note that the energy is shifted such that the Dirac point of the unstrained configuration sits at the Fermi level.\label{fig5}}
\end{center}
\end{figure}

\begin{figure}
\begin{center}
\includegraphics[height=0.95\linewidth,angle=-90]{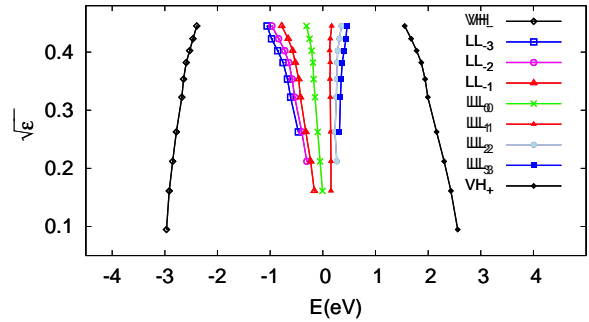}
\caption{(Color online) Plot of the energies corresponding to the
lower and higher van Hove singularities and the pseudo-Landau
levels, $n=\pm3,\pm2,\pm1,0$, as a function of $\sqrt{\epsilon}$
obtained from the peaks in the LDOS at the center of zig-zag
strained hexagon. \label{fig5p}}
\end{center}
\end{figure}
In order to investigate the effect of strain on the electronic
properties, we input the relaxed atomic positions obtained from our
 atomistic simulation into a real-space tight-binding model. Our main focus in this work is to find the LDOS
maps, which could be directly accessed by STM experiments. The LDOS
maps are obtained by expanding the Green's function at each atomic
position in terms of Chebyshev polynomials . The details of this
expansion can be found in our previous works \cite{covaci_superconducting_2011,covaci_efficient_2010,de_juan_space_2012,munoz_tight-binding_2012,*munoz_tight-binding_2013}.
Here we use both nearest neighbor and next-nearest neighbor contributions in the tight-binding Hamiltonian in order to account both for the strain
induced vector potential (nearest neighbor) and the strain induced scalar potential (next-nearest neighbor).

First, in Fig.~\ref{fig5}(a), we show the LDOS for an atom in the
A-sublattice (which are under stress at the three edges) in the
center of the hexagon for various applied force strengths.
 Several interesting effects can be observed. Because we also include next-nearest neighbor hopping amplitudes in the calculation,
 the Dirac point for the unstrained hexagon sits at a finite energy, $E_D=3\,t^\prime$, where $t^\prime$ is the next-nearest neighbor
 hopping amplitude. We therefore shift all the LDOS curves such that the Dirac point of the unstrained configuration
  sits at the Fermi level. We also observe an additional shift for the strained configurations because of the exponential suppression
  in $t^\prime$, which could be understood in terms of a strain induced scalar potential, which shifts the Dirac point downwards
  in energy~\cite{castro_neto_electronic_2009}. In addition, the reduction in the hopping amplitudes shift the van-Hove peaks, signaling also a change in the Fermi velocity.
 Another important effect, noted already in Ref.~[\onlinecite{guinea_energy_2009}],
  is the appearance of peaks in the LDOS for the strained zig-zag hexagon. These correspond to pseudo-Landau
  levels generated by the strong pseudo-magnetic field observed in the central region of the zig-zag hexagon. When compared to
  the regular Landau levels generated by real magnetic fields, one important difference can be
  seen: we find that the zeroth Landau level has a finite contribution to the LDOS only in one sub-lattice, i.e.
  the sublattice pertaining to the edge atoms under stress. For the non-zero pseudo-Landau levels the sublattice
  symmetry still holds. This can be seen in Fig.~\ref{fig5}(b) where we show the LDOS in the center of the strained hexagon (with $\epsilon=13\%$) for the A and B sub-lattice and compare
  them with the LDOS for the unstrained zig-zag case and the LDOS in the center of the strained armchair hexagon. Since the pseudo-magnetic field at the
  center of the armchair hexagon is small, the LDOS does not show pseudo-Landau levels, but shows a strain induced shift of both the Dirac point and the van Hove peaks.
\begin{figure}
\begin{center}
\includegraphics[height=\columnwidth,angle=-90]{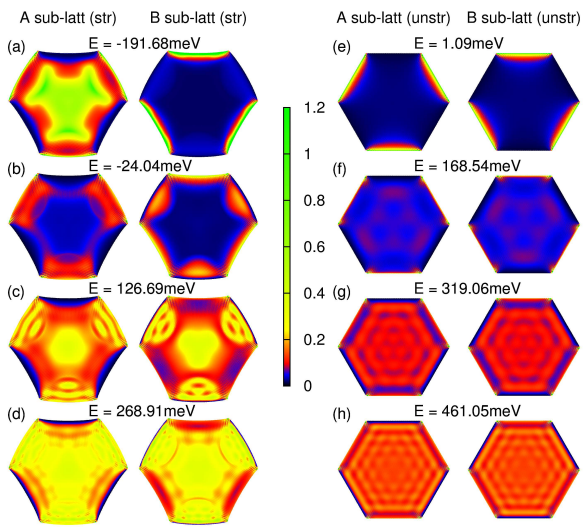}
\caption{(Color online) LDOS maps for a strained zig-zag hexagon with $\epsilon=13\%$, panels (a)-(d), and for an unstrained one, panels (e)-(h).
 Panel (a) corresponds to the zero pseudo-Landau level, while panels (c) and (d)
correspond to the first and second pseudo-Landau levels, respectively. \label{fig6}}
\end{center}
\end{figure}
The relativistic nature of the pseudo-Landau levels is clearly
apparent from Fig.~\ref{fig5}(c), where we plot the energy of the
pseudo-Landau levels for different strains as a function of
$sgn(n)\sqrt{nB}$. Note that we shift the pseudo-Landau levels such
that the zero-Landau level sits at zero energy. Surprisingly, the
relationship is linear. Note that the pseudo-magnetic field (also
shown in Fig.~\ref{fig4}(f)) was extracted from the energy gap
between the zero and
 first pseudo-Landau levels by using a constant Fermi velocity and
the known relationship: $E_{1}=\sqrt{2e\hbar v_F^2 B} \approx 30meV
\sqrt{B(Tesla)}$. Even though for low strains the pseudo-magnetic
field is proportional to $\epsilon$, the suppression of the Fermi
velocity (signaled by the shift of the van Hove peaks) will lead to
a non-linear dependence of the energy of the pseudo-Landau levels on
$\sqrt{\epsilon}$. This becomes more drastic at high strains, where
the pseudo-magnetic field does not follow anymore a linear
relationship with respect to
 $\epsilon$ (see Fig.~\ref{fig4}(f)). This can be seen more clearly in Fig.~\ref{fig5p}, where the energies of the van Hove peaks and of the pseudo-Landau levels
 are plotted as a function of $\sqrt{\epsilon}$. Although the spectrum of the unstrained graphene is not particle-hole symmetric, we fined that in the zig-zag strained graphene samples the higher-n pseudo-Landau levels are symmetric with respect to the zeroth pseudo-Landau level.

\begin{figure}
\begin{center}
\includegraphics[height=\columnwidth,angle=-90]{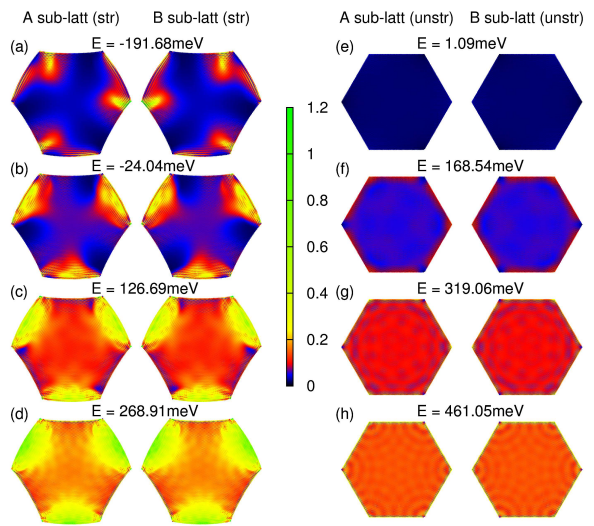}
\caption{(Color online) LDOS maps for a strained armchair hexagon with $\epsilon=13\%$, panels (a)-(d), and for an unstrained one, panels (e)-(h). \label{fig7}}
\end{center}
\end{figure}

Next we show in Fig.~\ref{fig6} the LDOS maps for various energies for a strained zig-zag hexagon with $\epsilon=13\%$. The contributions to each sub-lattice
are plotted separately for clarity. In panels (a)-(d) we show the
LDOS for the strained configuration,
 while in panels (e)-(h) the LDOS for the unstrained zig-zag hexagon is shown for corresponding energies. The energy of the LDOS map in panel (a)
 corresponds to the energy of the zero Landau level. We observe a large increase in the
 LDOS in sub-lattice A (which are under stress at the three edges) in the regions where the pseudo-magnetic
 field is large, together with a suppression of the LDOS in sub-lattice B. This effect shows the most important difference between pseudo-magnetic fields and
  real magnetic fields. Since a gap appears in the spectrum, and the time-reversal symmetry is not broken, this is a clear manifestation of a broken sub-lattice
  symmetry. If the hexagon would be pulled from the other three sides, the pseudo-magnetic field would change sign and the zero
  pseudo-Landau level will appear in the other sub-lattice.

The LDOS map shown in Fig.~\ref{fig6}(b) corresponds to an energy
close to the Dirac point of the unstrained configuration. The only
noticeable contributions come from the edge-like states, which are
again seen only in sub-lattice A. The energies of the LDOS maps
shown in Fig.~\ref{fig6}(c) and~\ref{fig6}(d) correspond to the
energies of the first and second pseudo-Landau levels, respectively.
In this case, we observe that the sub-lattice symmetry is preserved
in the central region of the hexagon. For these energies the real
and pseudo-Landau levels are similar. Interesting features appear
also at energies near the van Hove peaks, for which the LDOS maps
show resonant peaks at the center of the hexagon. At these energies peaks start developing, while there is no broken sub-lattice symmetry. It is important to note that the
low energy theory which relates the strain to a pseudo-vector potential is not valid near the van Hove peaks mainly because the dispersion is not linear anymore.
Movies of the LDOS maps for all energies  are presented as
Supplementary Materials \footnote{Movies of the LDOS maps for the
whole spectrum are presented as Supplementary Materials}.

Next we show for comparison in Fig.~\ref{fig6}(e)-(h) the LDOS maps for
the unstrained zig-zag hexagon. Due to the symmetries of the system,
i.e. $\pi/3$ rotational symmetry, the boundary conditions are such
that the wave-function is zero in sub-lattice A in three
non-consecutive sides and zero in sub-lattice B in the
 other three sides. At the Dirac point, panel (e), edge states with zero energy appear and are located in different sub-lattices on different sides of the hexagon.
  For all energies the LDOS map for the A sub-lattice is the same as the one for the B sub-lattice but rotated are $\pi/3$, as expected due to symmetry arguments.

The LDOS maps for the armchair hexagon are presented in
Fig.~\ref{fig7}, both for the strained, panels (a)-(d), and the
unstrained configurations, panels (e)-(h). The boundary condition
now sets the wave-function to zero in both sub-lattices at all
hexagon edges. The main symmetry, which is preserved also in the
strained configuration is the mirror symmetry plus a sub-lattice
exchange with respect to $x=0$. The effect of strain on the LDOS
maps is not so drastic as the one for the zig-zag hexagon. Besides
the overall shift in energy due to the appearance of the scalar
gauge field, the main
 modifications can be seen only at the edges of the hexagon where the pseudo-magnetic field is large. Since the field changes sign, we see localization of states
 at the Dirac point in both A and B sub-lattices.

\section{Summary}
In this paper we studied the effect of triaxial stress on the
electronic and structural properties of hexagonal flakes of graphene
with zig-zag and armchair edges. We combined molecular dynamics
simulations to obtain the relaxed atomic positions and the tight
binding method to describe the electronic properties.
 We found that lattice deformations under triaxial stress are well described by continuum
elasticity theory only for small strains ($\epsilon<15\%$) and only in
the central part of the sample. The pseudo gauge field was found to
be neither circular symmetric nor homogeneous in space, i.e. there
are modified triangular orbits for zig-zag flakes and none-orbital
vectors for armchair flakes when the deformed lattice is fully
relaxed. The corresponding pseudo-magnetic field is non-uniform over
the sample and exhibits three fold symmetry for a zig-zag flake and
a more complex variation for the armchair flake. Only for zig-zag
flakes we find that in the central region the pseudo-magnetic field
is large and while it varies significantly near the edges,
while for the armchair flakes the field is very small in the center and
oscillating near the edges. The local density of states are completely different at
the different sub-lattices and are mostly affected by the
pseudo-magnetic field distribution. In the zig-zag hexagon the
appearance of pseudo-Landau levels breaks the sub-lattice symmetry
in the zeroth pseudo-Landau level and all relaxed samples show  a
shift in the energy levels as compared to the undeformed case due to
the appearance of a strain induced scalar potential resulting from
the expansion of the graphene flake. We find that molecular dynamics
relaxation changes strongly the pseudo-magnetic field and the local
density of states as compared to the deformed none-relaxed samples
from
 elasticity theory. The latter shows that relaxation of
 the atomistic structure of the deformed graphene under constraints plays an important role in the electronic
 properties and that predictions of elasticity theory  applicable for continuum sheets should be modified, especially for large strains.

\acknowledgements
This work was supported by the
EU-Marie Curie IIF postdoctoral Fellowship/299855 (for M.N.-A.), the ESF
EuroGRAPHENE project CONGRAN, the Flemish Science Foundation
(FWO-Vl) and the Methusalem Funding of the Flemish government.

\appendix

\section{Effect of lattice relaxation}

\begin{figure}[b]
\begin{center}
\begin{minipage}{0.45\linewidth}
\includegraphics[width=\linewidth]{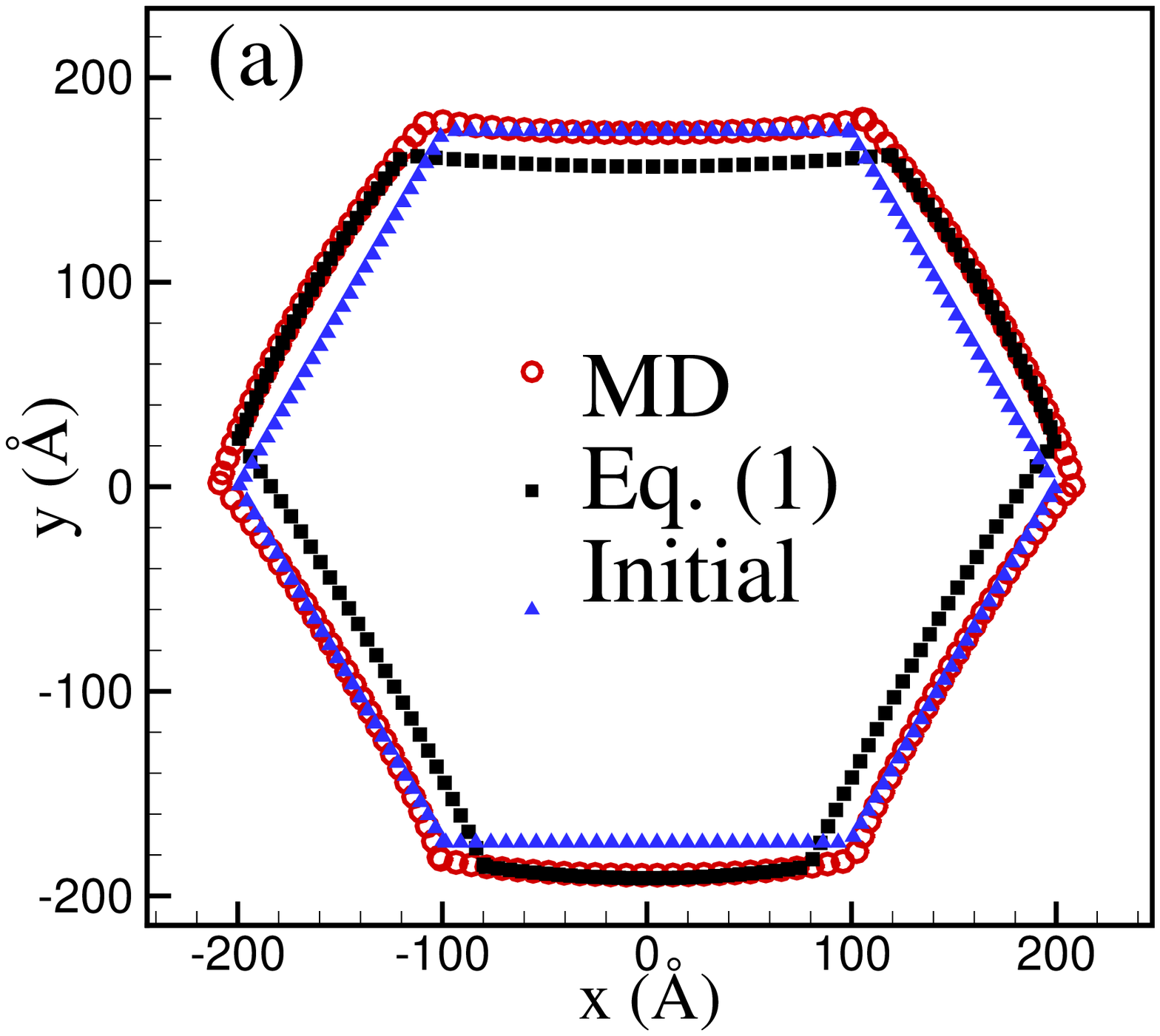}
\vspace{0.1mm}
\end{minipage}
\hspace{1mm}
\begin{minipage}{0.45\linewidth}
 \includegraphics[width=\linewidth]{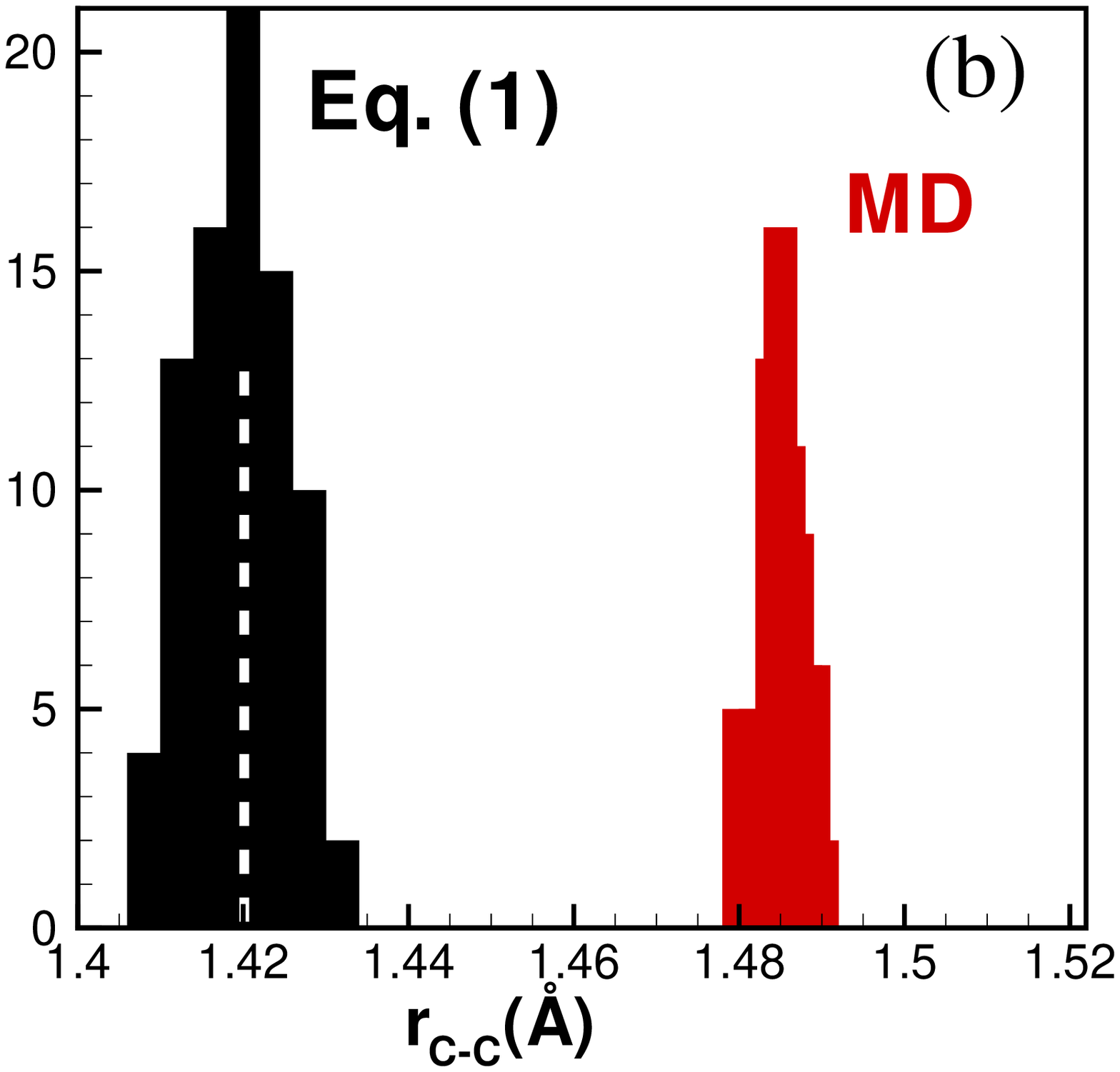}
\end{minipage}
\includegraphics[height=0.9\linewidth,angle=-90]{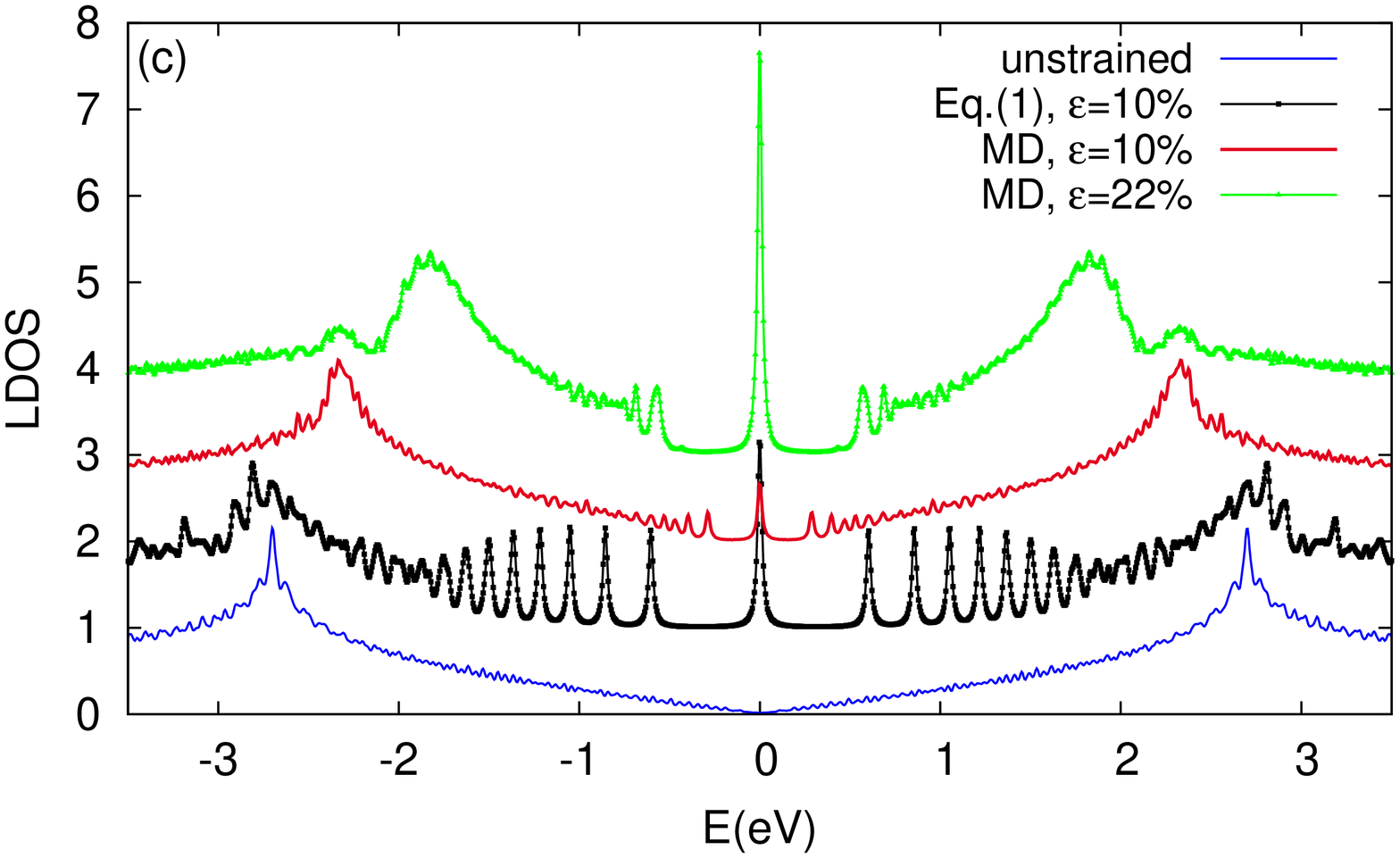}
\caption{(Color online) (a) The boundary of the deformed lattice
using Eq. (1) (black squares), MD relaxed flake (red circles) and
undeformed lattice (blue triangles) for a hexagonal flake subjected
to a strain of 10$\%$. (b) The C-C bond length histogram in the case
of MD relaxation and using Eq.~(1) which is the method used in
Ref.~[6]. (c)  Comparison between the central LDOS of the unstrained
zig-zag flake (blue curve), strained zig-zag flake for strain 22$\%$
(green curve) and 10$\%$ (red line) and that obtained by using
deformation based on Eq.~(1) (black curve). \label{figs1}}
\end{center}
\end{figure}
\begin{figure}[t]
\begin{center}
\includegraphics[height=0.95\linewidth,angle=-90]{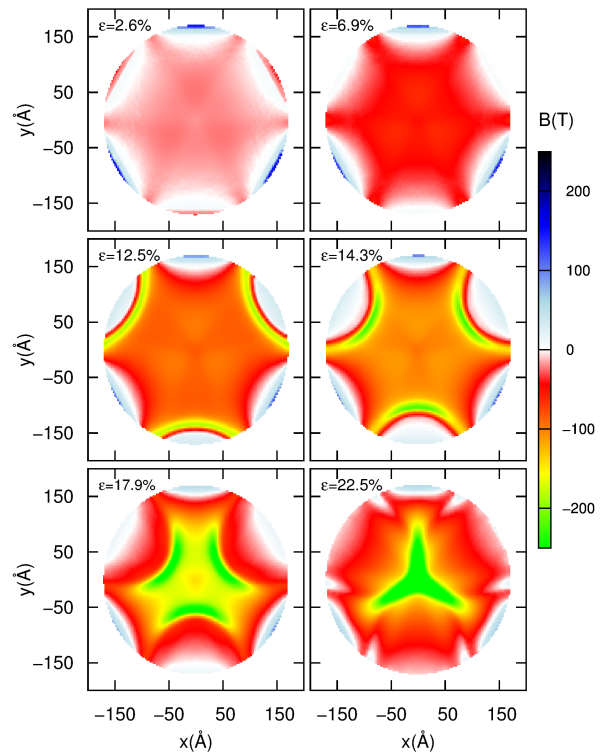}
\caption{(Color online) The effect of strain on the pseudo-magnetic
field in the central part of the zig-zag hexagonal graphene flake,
$|\textbf{r}|<17$\,nm. \label{figs2}}
\end{center}
\end{figure}

In this appendix we discuss the important effect of the  atomistic
relaxation of nano-scale samples using bond order force field on the
electronic properties of the studied hexagonal graphene flakes. One
of the simplest way to deform the perfect graphene lattice
non-uniformly (blue triangle in Fig.~\ref{figs1}) is by using directly Eq.~(1)
(here the case of triaxial strain). The boundary of the resulting
deformation for hexagonal graphene zig-zag flake subjected to 10$\%$
strain is shown in Fig. ~\ref{figs1}(a), see black square symbols.
The corresponding result obtained when using the MD relaxation is shown as open
red circles in Fig.~\ref{figs1}(a). As obviously from the figure, the MD relaxation
expands the area of flake, thus one expects longer C-C bond lengths. In order show the latter effect we plotted in Fig.
\ref{figs1}(b) the histogram of C-C bond lengths in the
central part ($|\textbf{r}|<$5\AA)~of hexagonal flakes for the two
before mentioned methods. The shorter the bond lengths, the more compact
the lattice and the higher the local stress in the central part will be,
which results in higher pseudo-magnetic field, see
blue triangles and red squares in Fig.~4(f) (see the results for
triaxial stressed unrelaxed hexagonal flake in
Ref.~\cite{masir_pseudo_2013} which are more than ten times higher
than those we report here). In order to reveal the large difference
between the pseudo-magnetic fields and the consequent effects on the
electronic spectrum  we depicted in Fig.~\ref{figs1}(c) the LDOS for
two relaxed systems by MD method for $\epsilon=10\%,22\%$, a
deformed system with $\epsilon=10\%$ (black curve) by using Eq.~(1)
and an undeformed hexagonal flake. Here we use only the
nearest-neighbor hopping amplitudes since the only effect of the
next-nearest neighbor term is to shift the Dirac point. The position
of the pseudo-Landau levels and the shift of the van Hove peaks are
not affected by the change in the next-nearest neighbor hopping. In
the case of an undeformed sample (ordinary hexagonal flake shown by
blue triangles in Fig.~\ref{figs1}(a)), we see that there is no zero
energy state at the Dirac point. By deforming the lattice following
Eq.~(1) with $\epsilon=10\%$ (hexagonal flake shown by black squares
in Fig. \ref{figs1}(a)) pseudo-Landau levels with
 energy separation proportional to $\sqrt{n}$ appear for both electrons and holes. The MD relaxation changes significantly the LDOS profiles
because of the mentioned lattice expansion. First, we observe that for the same deformation, $\epsilon=10\%$,
 the gap between the zero and first pseudo-Landau levels is much smaller, consistent with a much lower pseudo-magnetic
 field. We also find that fewer pseudo-Landau levels can be distinguished, even at high strains, e.g. $\epsilon=22\%$.
 This could be a consequence of the fact that at high strains the pseudo-magnetic field for the MD relaxed system is not
  constant, and therefore the high-$n$ pseudo-Landau levels are smeared out.

\section{Effect of strain on pseudo-magnetic field}

Here we show the effect of strain on the pseudo-magnetic field for
different strains in zig-zag hexagonal graphene flakes. The larger
the strain the larger the pseudo-magnetic field. In
Fig.~~\ref{figs2} we show the pseudo-magnetic field for strains in
the range of [2.6$\%$-22.5$\%$] in the central region, i.e
$|\textbf{r}|<17$\,nm.  It is seen that in the central portion of
the zig-zag hexagonal graphene flakes the pseudo-magnetic field is
not constant which is in contrast to the continuum elasticity theory
prediction~[6], and the  result obtains from
Eq.~(1),~$B=\frac{16\,C\,\hbar}{a_0\,e}$.

\section{Alternative methods for applying stress on the edges}

\begin{figure}[t]
\begin{center}
\includegraphics[height=0.99\linewidth,angle=-90]{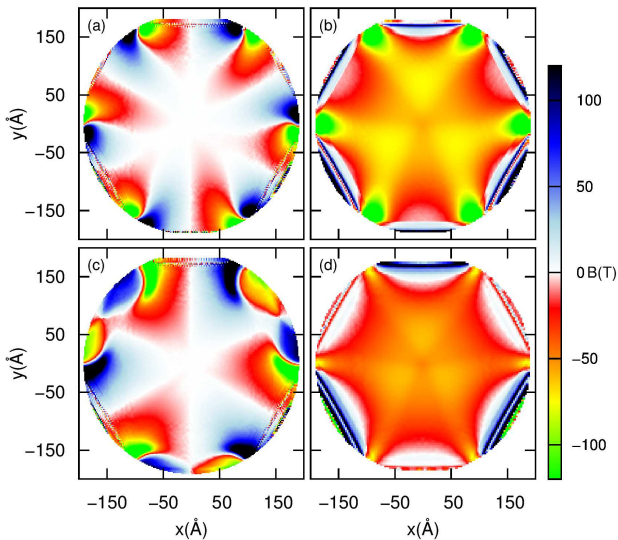}
\caption{(Color online) Pseudo-magnetic field in the central part of
the arm-chair (a),(c) and zig-zag (b),(d)  hexagonal graphene flake,
$|\textbf{r}|<17$\,nm where $\epsilon=13\%$. The fields are obtained
either by moving the edges (a),(b) or by applying a constant force
on the edges (c),(d). \label{figs3}}
\end{center}
\end{figure}

In addition to the method for applying the stress presented here,
one can fix the boundary atoms and shift them gradually in the
direction perpendicular to the edges. Since the boundary atoms are
not allowed to be relaxed during the simulation, the inter-atomic
distances at the edge remain constant giving high stress at the
corners, which consequently affects the stress distribution through
the system. The latter effect modifies both the pseudo-magnetic
field and the LDOS. We found that in case of such a fixed boundary
condition the pseudo-magnetic field is higher than
the one found from the constant applied force method, as
shown in Figs.~\ref{figs3}(a-d). Moreover, the pseudo-magnetic is
also more inhomogeneous. The recent experimental realization of
triaxial strain in molecular graphene~\cite{gomes_designer_2012}
confirms that the method used in the present work is realistic. The
constant motion of the edges was recently used by Z. Qi \emph{et
al.} in order to study the resonant tunneling in hexagonal
graphene~\cite{qi_resonant_2013} quantum dots.  This should be
modified in order to have the relevant order of magnitude of the
pseudo-magnetic field and consequently the correct position of the
Landau levels.

\bibliographystyle{apsrev4-1}
%
\end{document}